\documentclass[showpacs,preprintnumbers,amsmath,two column]{revtex4}
\usepackage{graphicx}
\usepackage{dcolumn}
\usepackage{bm}

\begin{document}

\title{ Generalized rotating-wave approximation to biased  qubit-oscillator systems}
\author{Yu-Yu Zhang$^{1}$, Qing-Hu Chen$^{2,3}$, and Yang Zhao$^{4}$  }
\address{
$^{1}$Center for Modern Physics, Chongqing University, Chongqing
400044, P. R.  China\\
$^{2}$Department of Physics, Zhejiang University, Hangzhou 310027,
P. R. China\\
$^{3}$Center for Statistical and Theoretical Condensed Matter
Physics, Zhejiang Normal University, Jinhua 321004, P. R. China\\
$^{4}$Division of Materials Science, Nanyang Technological University, Singapore 639798, Singapore
}

\date{\today}

\begin{abstract}
The generalized rotating-wave approximation with counter-rotating
interactions has been applied to
a biased qubit-oscillator system. Analytical expressions are explicitly given
for all eigenvalues and eigenstates. For a flux qubit coupled to superconducting oscillators,
spectra calculated by our approach are in excellent agreement with experiment.
Calculated energy levels for a variety of biases also agree well
with those obtained via exact diagonalization for a wide range of coupling strengths.
Dynamics of the qubit has also been examined, and results lend further support to the validity of the
analytical approximation employed here. Our approach
can be readily implemented and
applied to superconducting
qubit-oscillator experiments conducted currently and in the near future
with a biased qubit and for all accessible coupling strengths.

\end{abstract}

\maketitle

\section{Introduction}
The combination of a two-level system (or a qubit) and a harmonic oscillator
has found myriad interesting
applications in quantum systems
ranging from two-level atoms coupled to optical or microwave
cavities~\cite{Scully,Buluta} to superconducting qubits coupled to
superconducting resonators~\cite{schuster,Wallraff,chio,you,nation}.
In early work on cavity quantum electrodynamics (QED), the
qubit-oscillator coupling strength $g$ achieved was much smaller
than the cavity transition frequency $\omega$, i.e., $g/\omega\sim0.001$.
Experiments can therefore be well described by the Jaynes-Cumming model
with the rotating-wave approximation (RWA)~\cite{jaynes}.

In recent circuit QED setups, where artificial superconducting two-level
atoms are coupled to on-chip cavities, the exploration of quantum
physics has greatly evolved in the ultrastrong coupling regime,
where the atom-cavity coupling strength is comparable to the cavity
transition frequency,
$g/\omega\sim0.1$~\cite{Wallraff,chio,Niemczyk,pfd,fedorov}. It is
evident for the breakdown of the RWA and the counter-rotating terms
are expected to take effect. There have been numerous theoretical
studies on the qubit-oscillator system finding new phenomena in
the ulrastrong coupling regime ~\cite{Ashhab,grifoni,irish,feranchuk,amniat} and deep
strong coupling regime with $g/\omega>1$~\cite{casanova,zhang}.
However,
many theories are derived for an unbiased qubit, or in the terminology
of cavity and circuit QED, for a qubit operated at the degeneracy point
or the sweet spot. While the unbiased qubit is often encountered for real atoms
in cavity QED, it is quite straightforward to vary the static bias of superconducting
qubits by adjusting an external control parameter such as the gate voltage applied the magnetic
flux acting on a Josephson junction~\cite{Niemczyk,pfd,fedorov}.
Therefore, it is necessary to develop theories which adequately treat
the biased qubit-oscillator system. Taking the
qubit's bias into account, Grifoni et al.~\cite{grifoni} formulated the
Van Vleck perturbation (VVP) theory beyond the RWA to treat
analytically a two-level system coupled to a harmonic oscillator.
Unfortunately, it gives unphysical energy-level crossings in the weak
coupling regime for positive detuning.
An adiabatic approximation was proposed by Nori et al.~\cite{Ashhab} for a biased system
in various parameter regimes.
Existing approaches to describe the behavior of the biased system are often suited for a particular circumstance, despite that
many analytical methods have been proposed.
An efficient, accurate treatment of the biased qubit-oscillator system in all parameter regimes remains elusive.

In this work a generalized rotating-wave
approximation (GRWA) is proposed for the biased qubit-oscillator system in the
ulstrastrong coupling regime, extending the pioneering work of Irish on unbiased system~\cite{irish}, an approach that
we shall call the biased generalized rotating-wave approximation (BGRWA).
Our analytical approach takes into account the effect of qubit-oscillator
counter-rotating terms, while the renormalized Hamiltonian including energy-conserving terms
retains the mathematical simplicity of the usual RWA.
This easily-implemented approach gives simple analytical expressions for eigenvalues and eigenvectors for the ground and low-lying excited states, and is applicable to a wide range of the coupling parameters.
In the limit of zero bias, we recover the early results of Irish~\cite{irish} which were obtained
for an unbiased qubit using GRWA.  By parameter-fitting the
circuit QED experiment, an analytical experssion is obtained
for the energy spectrum
in the ulstrastrong coupling regime.
Validity of our approach is discussed by comparing with the VVP method as well as numerical exact diagonalization.
Furthermore, we also study the
qubit dynamics in the ultrastrong coupling regime to confirm the
effectiveness of the BGRWA.

The paper is outlined as follows. In Sec.~II, we derive expressions for the eigenenergies and eigenstates
of a biased qubit-oscillator system using BGRWA. Analytical expressions for the spectrum is also
given by fitting the circuit QED experiment. In Sec.~III, we discuss the qubit dynamics in the finite bias case.
Finally, a brief summary is given in Sec.~IV.

\section{analytical solution}

The Hamiltonian of the qubit can be written as
$H_{q}=-(\varepsilon\sigma_x+\Delta\sigma_z)/2$, where $\sigma_x$
and $\sigma_z$ are Pauli matrices; $\Delta$ is the tunneling
parameter between the upper level $|+z\rangle$ and the lower level
$|-z\rangle$ in the basis of $\sigma_z$; $\varepsilon$ is the
magnetic energy bias related to the circulating current in the qubit
loop and the applied magnetic flux~\cite{pfd,fedorov}. In the weak
coupling regime, where the interaction strength $g$ exceeds the
cavity and qubit loss rates, the RWA can be applied and the system
can be described by the Jaynes-Cummings-type Hamiltonian for zero
bias as
\begin{equation}\label{RWA}
H=\frac \Delta 2\sigma _z+\omega a^{\dagger
}a+g(a^{\dagger }\sigma_-+a\sigma_+),
\end{equation}
where $a^\dagger$ and $a$ are the creation and annihilation
operators for the oscillator, and we have set $\left( \hbar \right) $ to unity.
The Jaynes-Cummings-type Hamiltonian (\ref{RWA}) can be solved analytically in a
closed form in the basis $|n\rangle|+z\rangle$ and
$|n+1\rangle|-z\rangle$, where the qubit states $|\pm z\rangle$ are
eigenstates of $\sigma_z$, and the oscillator states $|n\rangle$
($n=0,1,2,...$) are Fock states. The ground state obtained is
$|\psi_g\rangle=|0\rangle|-z\rangle$ for weak coupling. However, current
experimental advances draw our attention to the ultrastrong coupling
regime, where $g$ approaches to the qubit or oscillator frequencies,
and the RWA no longer holds~\cite{Niemczyk,pfd,fedorov}. Thus, the
qubit-oscillator counter-rotating interaction $a^{\dagger
}\sigma_++a\sigma_-$ needs to be taken into account.

Under a rotation around the $y$ axis with the angle $\pi/2$, the
Hamiltonian of the qubit-oscillator system including  the
counter-rotating terms reads
\begin{equation}\label{hamiltonian}
H=-\frac \Delta 2\sigma _x-\frac \varepsilon 2\sigma _z+\omega a^{\dagger
}a+g(a^{\dagger }+a)\sigma _z.
\end{equation}
Making use of a unitary transformation
$
U=\exp \left[ -\frac g\omega \sigma _z\left( a-a^{\dagger }\right) \right],
$
we can obtain a transformed Hamiltonian $H^{'}=U^{\dagger}HU=H_0+H_1$, consisting of
\begin{eqnarray}
H_0 &=&\omega a^{\dagger }a-g^2/\omega
-\frac \varepsilon 2\sigma _z, \\
H_1 &=&-\frac \Delta 2 \{\sigma _x\cosh[\frac{2g}{\omega} \left(
a^{\dagger }-a\right)]+i\sigma _y\sinh[\frac{2g}{\omega} \left(
a^{\dagger }-a\right)]\}. \nonumber\\
\end{eqnarray}

Recently, much theoretical attention has been devoted to the qubit-oscillator
system using a variety of
transformations~\cite{grifoni,irish,yu,zheng,talab}.
In particular, Irish has presented a
generalized version of the RWA by performing a simple basis change prior to eliminating the counter-rotating terms \cite{irish}.
This gives rise to a significantly more accurate expression for the energy levels of the system for all values of the coupling strength.
We now extend
the generalized RWA by Irish to the biased
qubit-oscillator system. The simplicity of the approximation is
based on its close connection to the standard RWA. Consequently, the
terms retained in $H_1$ correspond to the energy-conserving
one-excitation terms, just as in the standard RWA. When $\cosh\left[
\frac{2g}\omega \left( a^{\dagger }-a\right) \right]$ is expanded as
$1+\frac{1}{2!}[\frac{2g}{\omega} \left( a^{\dagger
}-a\right)]^2+\frac{1}{4!}[\frac{2g}{\omega} \left( a^{\dagger
}-a\right)]^4+..., $ it is performed by keeping the terms containing
the number operator $a^{\dagger}a=n$ with the coefficient $G_0\left(
n\right)$
\begin{eqnarray}
G_0(n)&&=\langle n|\cosh[ \frac{2g}\omega( a^{\dagger }-a)]|n\rangle\nonumber\\
&&= \exp(-2g^2/\omega^2)L_n(4g^2/\omega^2),
\end{eqnarray}
where $L_n$ are the Laguerre polynomials. Higher-order excitation terms
such as $a^{\dagger 2}$, $a^2$,..., which are accounted for
multi-photon process, are neglected within this  approximation.
Similarly, by expanding $ \sinh[\frac{2g}{\omega} \left( a^{\dagger
}-a\right)]=\frac{2g}{\omega} \left( a^{\dagger
}-a\right)+\frac{1}{3!}[\frac{2g}{\omega} \left( a^{\dagger
}-a\right)]^3+\frac{1}{5!}[\frac{2g}{\omega} \left( a^{\dagger
}-a\right)]^5+..., $ the one-excitation terms are kept as $F_1\left(
n\right) a^{\dagger }-aF_1\left(n\right)$ with the coefficient
$F_1\left( n\right)$ to be determined. Since the terms $aF_1\left(
n\right) $ and $F_1\left( n\right) a^{\dagger }$ involve creating
and eliminating a single photon of the oscillator, it can be
evaluated as
\begin{eqnarray}
F_1\left( n\right)
&&=\frac 1 {\sqrt{n+1}}\left\langle n+1\right| \sinh \left[ \frac{2g}
\omega \left( a^{\dagger }-a\right) \right] \left| n\right\rangle \nonumber\\
&&=\frac{2g}{\omega(n+1)}e^{-2g^2/\omega^2}L_n^1(4g^2/\omega^2).
\end{eqnarray}
Since the higher-order terms of $H_1$ are discarded, we can construct an effective Hamiltonian $H^{'}= H_0^{'}+H_1^{'}$ with
\begin{eqnarray}\label{transformed hamiltonian}
H_0^{'}&=&\omega a^{\dagger }a-g^2/\omega -\frac{\Delta \eta }2\sigma _x-\frac
\varepsilon 2\sigma _z, \nonumber\\
H_1^{'} &=&-\frac \Delta 2[G_0(n)-\eta ]\sigma _x-i\frac \Delta 2\sigma _yF_1(n)(a^{\dagger }-a),\nonumber\\
\end{eqnarray}
where the parameter $\eta$ is defined as $\eta =G_0(0)$.

As the qubit and oscillator are decoupled in $H_0^{'}$ and its
qubit part can be diagonalized by a second unitary transformation $
S=\left(
\begin{array}{ll}
u & v \\
v & -u
\end{array}
\right)
$
with $u=\frac 1{\sqrt{2}}\sqrt{1-\frac \varepsilon y}$, $v=\frac 1{\sqrt{2}}%
\sqrt{1+\frac \varepsilon y}$, and $y=\sqrt{\varepsilon ^2+\Delta ^2\eta ^2}$.
The diagonalized $H_0^{'}$ takes the form
\begin{eqnarray}
\tilde{H_0} &=&S^{+}H_0^{'}S=\omega a^{\dagger }a-g^2/\omega +\frac 12\sqrt{%
\varepsilon ^2+\Delta ^2\eta ^2}\sigma _z, \nonumber\\
\end{eqnarray}
where the tunneling parameter is renormalized by $\sqrt{%
\varepsilon ^2+\Delta ^2\eta ^2}/2$. And the $H_1^{'}$ is transformed into
\begin{eqnarray}
\tilde{H_1} &=&S^{+}H_1^{'}S \nonumber\\
&=&\frac{\Delta ^2\eta[G_0(n)-\eta ] }{2\sqrt{\varepsilon ^2+\Delta ^2\eta ^2}}\sigma _z
-\frac \Delta 2F_1(n)(\sigma _+-\sigma_-)(a^{\dagger }-a)\nonumber\\
&-&\frac{\Delta \varepsilon[G_0(n)-\eta ] }{2\sqrt{\varepsilon ^2+\Delta ^2\eta ^2}}\sigma _x.
\label{eq9}
\end{eqnarray}

In order to cast the second term in Eq.~(\ref{eq9}), representing the qubit-socillator interactions in $\tilde{H_1}$,
into the same form as the ordinary RWA term in Eq.~(\ref{RWA}),  the
Hamiltonian under BGRWA can be approximated by the form
\begin{eqnarray}\label{hamiltonian new}
H_{\rm BGRWA}=\omega a^{\dagger }a-g^2/\omega +\varepsilon(n)\sigma _z
+R_r(a^{\dagger }\sigma _{-}+a\sigma _{+}),\nonumber\\
\end{eqnarray}
where the tunneling parameter $\varepsilon(n)$ is renormalized to
\begin{equation}
\varepsilon(n)=\frac{\varepsilon ^2+\Delta ^2\eta
G_0(n)}{2\sqrt{\varepsilon ^2+\Delta ^2\eta ^2}},
\end{equation}
and the effective coupling strength is
$R_r=\Delta F(n)/2$, which depends on the parameters $\Delta$ and
$g$. The Hamiltonian after the transformation retaining the
mathematical structure of the ordinary RWA contains the
counter-rotating terms, which play an important role in the
ultrastrong coupling regime.

Our aim is to extend the GRWA derivation to qubit-oscillator systems
with a finite bias.
Similar to the GRWA employed by Irish~\cite{irish}, only zero- and one-excitation terms are kept in the transformed Hamiltonian in terms of $G_{0}(n)$ and $F_{1}(n)$. Unlike the GRWA for zero bias, we take into account the static bias of the qubit while adjusting the renormalized tunneling parameter $\varepsilon(n) $, a term also present in the transformation of the biased spin-boson model by Gan and Zheng~\cite{zheng}.
The effective Hamiltonian (\ref{hamiltonian new}) with the counter-rotating interactions
contains the energy-conserving term $R_{r}(a^{+}\sigma_{-}+a\sigma_{+})$, which is identical in form to
the corresponding term in the usual RWA Hamiltonian (\ref{RWA}).
A simplified expression for a biased qubit system, our approximation
is expected to extend the range of validity to the ulstrastrong coupling regime for
qubit-oscillator systems with a finite bias.

One can easily diagonalize the Hamiltonian (\ref{hamiltonian new}) in the basis of $|+z,n\rangle
$ and $|-z,n+1\rangle $
\begin{widetext}
\begin{equation}
H_{\rm BGRWA}=\left(
\begin{array}{ll}
\omega n-g^2/\omega +\varepsilon(n) & R_r(n)\sqrt{n+1} \\
R_r(n)\sqrt{n+1} & \omega (n+1)-g^2/\omega -\varepsilon(n)
\end{array}
\right).
\end{equation}
\end{widetext}
It is straightforward to obtain the eigenvalues
\begin{widetext}
\begin{eqnarray}\label{eigenvalues}
E_n^{\pm ,{\rm BGRWA}}
&=&\omega (n+\frac 12)-g^2/\omega +\frac{\Delta ^2\eta }{4\sqrt{\varepsilon
^2+\Delta ^2\eta ^2}}e^{-2g^2/\omega ^2}[L_n(4g^2/\omega
^2)-L_{n+1}(4g^2/\omega ^2)] \nonumber \\
&&\pm \{[\frac \omega 2-\frac{2\varepsilon ^2+\Delta ^2\eta e^{-2g^2/\omega
^2}[L_n(4g^2/\omega ^2)+L_{n+1}(4g^2/\omega ^2)]}{4\sqrt{\varepsilon
^2+\Delta ^2\eta ^2}}]^2+\frac{g^2\Delta ^2e^{-4g^2/\omega ^2}}{\omega
^2(n+1)}[L_n^1(4g^2/\omega ^2)]^2\}^{1/2},\nonumber\\
\end{eqnarray}
\end{widetext}
and the corresponding eigenfunctions
\begin{eqnarray}
|\varphi _n^{+}\rangle &=&\cos \frac \theta 2|n\rangle |+z\rangle +\sin
\frac \theta 2|n+1\rangle |-z\rangle , \\
|\varphi _n^{-}\rangle &=&\sin \frac \theta 2|n\rangle |+z\rangle -\cos
\frac \theta 2|n+1\rangle |-z\rangle ,
\end{eqnarray}
where
\begin{eqnarray}
\theta & =&\arccos \frac \delta {\sqrt{\delta ^2+4R_r^2(n+1)}}\nonumber \\
\delta&=&\frac{2\varepsilon ^2+\Delta ^2\eta [L_n(4g^2/\omega ^2)+L_{n+1}(4g^2/\omega ^2)]}{%
2\sqrt{\varepsilon ^2+\Delta ^2\eta ^2}}-\omega \nonumber
\end{eqnarray}
In the case of $\varepsilon =0$, the eigenvalues in Eq.~(\ref{eigenvalues}) are reduced to the GRWA form~\cite{irish}.
The energy for the ground-state $|-z,0\rangle $ is
\begin{equation}\label{groud-state energy}
E_g^{\rm BGRWA}=-\frac 12\sqrt{\varepsilon ^2+\Delta ^2\eta^2}-g^2/\omega .
\end{equation}

\begin{figure}[tbp]
\includegraphics[trim=55 20 30 20,scale=0.7]{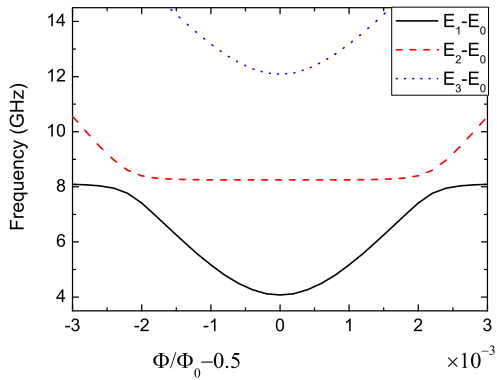}
\caption{Spectrum of the qubit coupled to an oscillator obtained
from Eqs.~(\ref{eigenvalues})  and (\ref{groud-state energy}) using
the experiment parameters in Ref~\cite{pfd}.} \label{spectrum}
\end{figure}

Thanks to the recent advances in experiment,
new spectral observations on qubit-oscillator systems are made available in the ultrastrong coupling
regime~\cite{pfd}, which were fitted by exact
diagonalization using the Fock basis \cite{pfd} and the
coherent-state basis \cite{lei}. In the setup of a flux
qubit coupled to an superconducting oscillator, the bias parameter
$\varepsilon=2I_p(\Phi-\Phi_0/2)$ with $I_p$ the persistent current
in the qubit loop, $\Phi$ an externally applied magnetic flux, and
$\Phi_0=\hbar/2e$ the flux quantum.  Fig.~\ref{spectrum} shows the
spectrum of the system using Eqs.~(\ref{eigenvalues}) and
(\ref{groud-state energy}), with fitted parameters of the
experimental results $g/2\pi=0.82~\rm{GHz}$, $\omega/2\pi=8.13~\rm{GHz}$,
$\Delta=4.25~\rm{GHz}$ and $I_p=510~nA$. No substantial difference is found
between our results and experiment ones as shown in Fig.~3 of Ref.~\cite{pfd}.
This demonstrates the great potential of our BGRWA approach to be applied in
future experiment as increasingly larger coupling strengths become accessible.

\begin{center}
\begin{figure*}
\includegraphics[width=\textwidth]{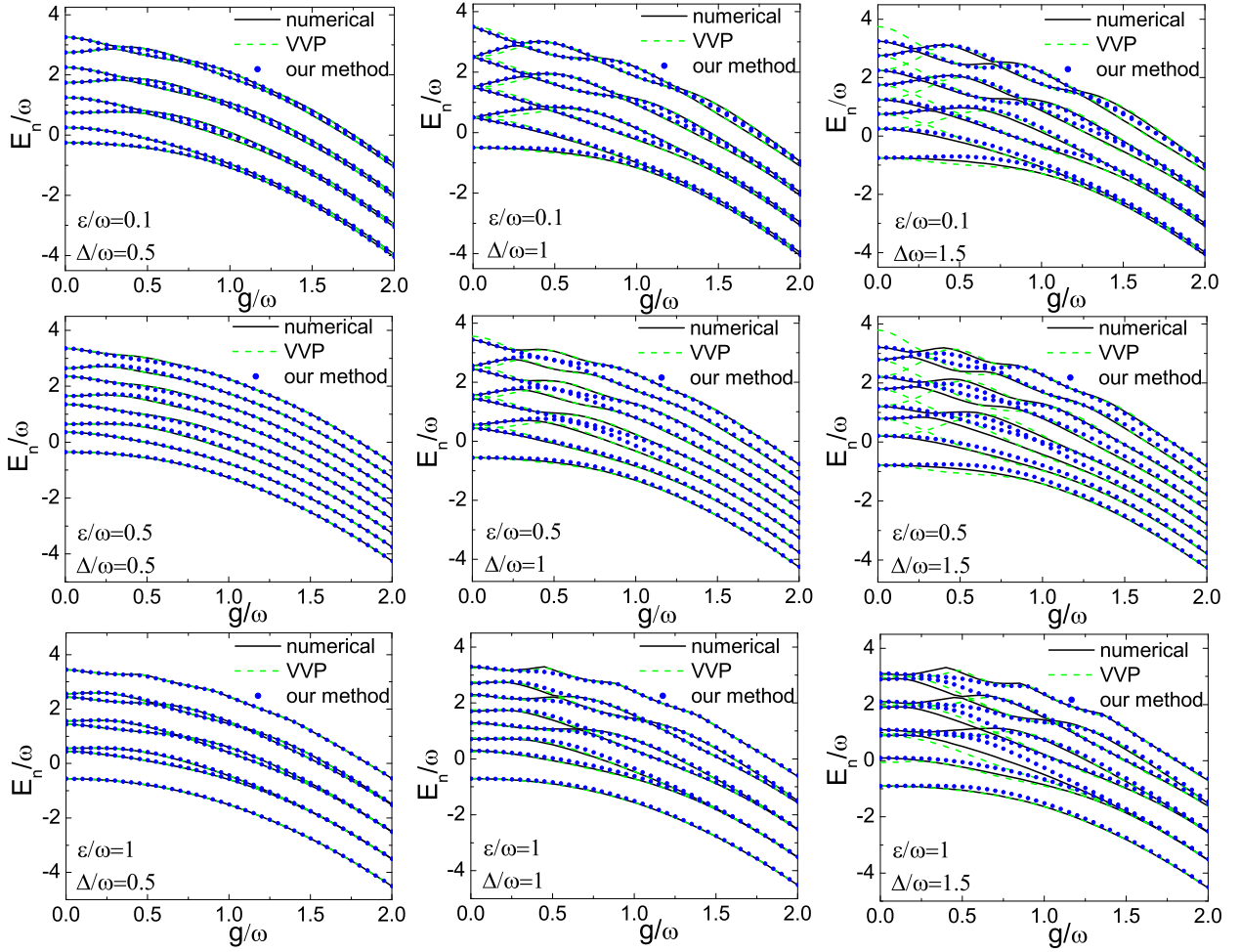}
\caption{Energy levels $E_n$ as a function of coupling strength $g$ for different bias $\varepsilon=0.1$, $0.5$ and $1$ (from top to bottom) with the tunneling parameter $\Delta=0.5$, $1$ and $1.5$. We compare the eigenvalues in Eq.~(\ref{eigenvalues}) (solid dot) with those obtained by the numerical exact diagonalization method (solid line) and VVP eigenvalues in Eq.~(\ref{vvp equation})(dashed line). We set $\omega=1$.}\label{epsilon1}
\end{figure*}
\end{center}

To the best of our knowledge,
Eqs.~(\ref{eigenvalues})-(\ref{groud-state energy}), obtained using
our BGRWA approach, is the simplest among all existing analytical
expressions. Provided that its validity is verified in general,
the BGRWA approach is a potentially effective tool in the study of
the superconducting qubit-oscillators where the biased parameter
can be adjusted externally.  To this end, we
present here a detailed discussion on the energy spectrum of the biased qubit-oscillator system.
First, we consider
eigenvalues obtained by the VVP method,
which can be written as
~\cite{grifoni}
\begin{widetext}
\begin{eqnarray}\label{vvp equation}
E_m^{\pm}&=&(m+\frac{l}{2})\omega-\frac{g^{2}}{\omega}+\frac{1}{2}\sum_{k=0,k\neq m\pm l}(\frac{D_{mk}^2}{\varepsilon+(m-k)\omega}-\frac{D_{nk}^2}{\varepsilon+(k-n)\omega})\nonumber\\
&\pm& \frac{1}{2}\sqrt{[\varepsilon-l\omega+\Delta^2\sum_{k=0,k\neq m\pm l}(\frac{D_{mk}^2}{\varepsilon+(m-k)\omega}+\frac{D_{nk}^2}{\varepsilon+(k-n)\omega})]^2+4D_{mn}^2},\nonumber\\
( n&=&m+l,l\geq0)
\end{eqnarray}
\end{widetext}
where $D_{mn}=\frac{\Delta}{2}(-1)^m(2g)^{n-m}e^{-2g^2}\sqrt{\frac{m!}{n!}}L_m^{n-m}(4g^2/\omega^2)$. The $m$th eigenvalues $E_m^{\pm}$ is a mixture of the oscillator levels $m$ and $l$.  Note the ambiguity of the value of $l$, which is selected to give better results. The VVP method works well for large values of the bias $\varepsilon$ and strong qubit-oscillator coupling.  In comparison, our analytical expression of eigenvalues as given in Eq.~(\ref{eigenvalues}) can be more easily implemented.
Below we will give a detailed comparison for various values of the coupling strength $g$ and detuning parameter $\Delta/\omega$.

Fig.~\ref{epsilon1} displays the first eight energy levels as a function of the
coupling strength $g$ for various values of the bias
$\varepsilon$ and the tunneling parameter $\Delta$. Here we set
$\omega=1$. For negative detuning $\Delta=0.5$, our analytical
approach and the VVP method are in good agreement with the numerical exact-diagonalization results
from the weak coupling regime to the strong coupling regime for $\varepsilon=0.1$, $0.5$ and $1$, as shown in
the left column of Fig.~\ref{epsilon1}. At the resonance $\Delta=1$ (middle
column), our analytical solutions agree well with the numerical
results for $g<0.5$, a coupling strength range that is either currently accessible
($g<0.12$) \cite{Niemczyk} or will be made
accessible in the near future.
In this interesting coupling regime of $g<0.5$, the VVP results deviate
considerably from the numerical ones.
In the intermediate coupling regime ($0.5<g<1$), there is a noticeable difference between results from our method and those from
the exact diagonalization due to the dominant influence of
the higher-order terms neglected in the transformed hamiltonian in
Eq.~(\ref{transformed hamiltonian}) accompanied by more photon excitations.
In the case of positive detuning $\Delta=1.5$, substantial improvements of our approach over the VVP method can be seen in the weak coupling regime, as shown in the right column of Fig.~\ref{epsilon1}. Especially, for $\varepsilon=0.1$
and $0.5$, the VVP results in the weak coupling regime are
qualitatively incorrect with an unphysical crossing. In comparison, our
BGRWA results remain in agreement with the numerically exact ones
one. Therefore, the BGRWA approach, which takes into
account the effect of counter-rotating terms, provides an efficient, yet accurate analytical expressions to the energy spectrum of the biased qubit-oscillator system.

\begin{figure*}[tbp]
\includegraphics[width=\textwidth]{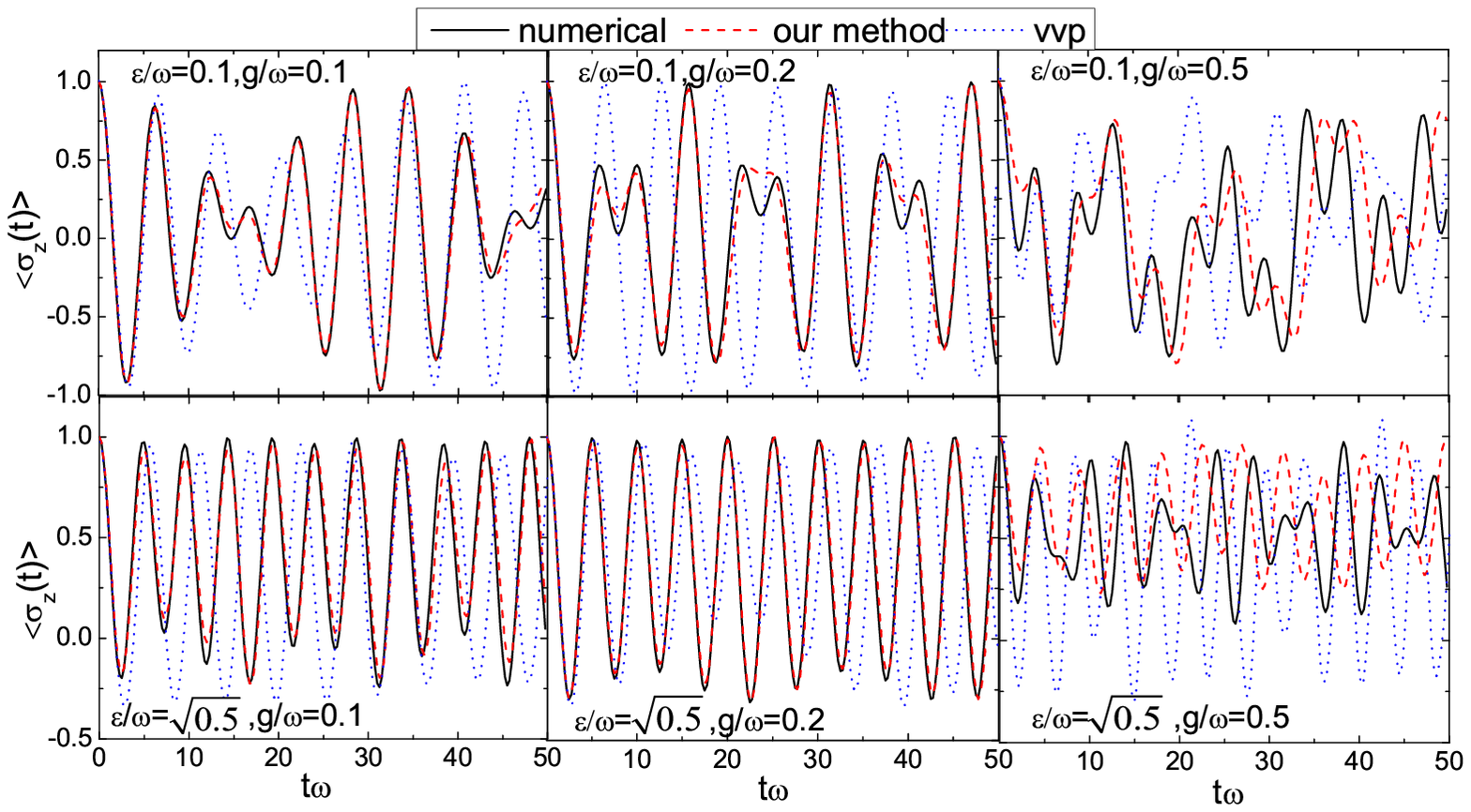}
\caption{Time evolution of $\sigma_z(t)$ for $\varepsilon=0.1$ (upper panel) and $\sqrt{0.5}$ (low panel) for different coupling strength $g=0.1,0.2,0.5$ on resonance. The dashed line is our BGRWA calculation. The solid and the dotted line are the numerically exact result and the VVP one, respectively. The evolution starts with a vacuum oscillator state $|0\rangle$ and an excited spin $|e\rangle$.  }
\label{dynamics}
\end{figure*}

\section{dynamics of the qubit}

In the original Hamiltonian (\ref{hamiltonian}) with counter-rotating terms, the excited
wave functions without RWA can be obtained using a unitary
transformation $|\Psi _n^{\pm}\rangle=U^{+}S^{+}|\varphi
_n^{\pm}\rangle$:
\begin{eqnarray}\label{wavefunction}
|\Psi _n^{+}\rangle^{\rm BGRWA}
&=&(u \cos \frac \theta 2 |n\rangle_{\frac{g}{\omega}}+v \sin \frac \theta 2|n+1\rangle_{\frac{g}{\omega}})|+x\rangle\nonumber\\
&+&(v \cos \frac \theta 2 |n\rangle_{\frac{-g}{\omega}}-u \sin \frac \theta 2|n+1\rangle_{\frac{-g}{\omega}})|-x\rangle,\nonumber\\
|\Psi _n^{-}\rangle^{\rm BGRWA}
&=&(u \sin \frac \theta 2 |n\rangle_{\frac{g}{\omega}}-v \cos \frac \theta 2|n+1\rangle_{\frac{g}{\omega}})|+x\rangle\nonumber\\
&+&(v \sin \frac \theta 2 |n\rangle_{\frac{-g}{\omega}}-u \cos \frac \theta 2|n+1\rangle_{\frac{-g}{\omega}})|-x\rangle,\nonumber\\
\end{eqnarray}
where the qubit states $|\pm x\rangle$ are eigenstates of $\sigma_x$, and the oscillator states $|n\rangle _{_{\pm g/\omega }}=e^{\mp g/\omega (a-a^{\dagger
})}|n\rangle $ are displaced Fock states, or the so-called coherent states.
The ground-state wave function is obtained by
\begin{eqnarray}\label{ground state}
|\Psi _g\rangle^{\rm BGRWA}
=v|e\rangle e^{-g/\omega (a-a^{\dagger
})}|0\rangle-u|g\rangle e^{g/\omega (a-a^{\dagger
})}|0\rangle.\nonumber\\
\end{eqnarray}

Next we examine the time evolution of $\sigma_z(t)$ to further demontrate
the validity of our analytical approach. The initial state is
assumed as $|\varphi(0)\rangle=|\uparrow\rangle|0\rangle$. Using the
eigenvectors $\{|\Psi_n\rangle^{\rm BGRWA}\}$ and eigenevalues
$\{E_n^{\rm BGRWA}\}$, the dynamical wave function of the Hamiltonian (\ref{hamiltonian}) without RWA can be expressed as
\begin{eqnarray}
|\varphi(t)\rangle &&=e^{-iHt}|\varphi(0)\rangle \nonumber\\
&&=\sum_n e^{-itE_n^{\rm BGRWA}}|\Psi _n\rangle^{\rm BGRWA}\langle\Psi _n|\varphi(0)\rangle.\nonumber\\
\end{eqnarray}
Using our approach without RWA for the biased system,
$\langle\sigma_z(t)\rangle=\langle\varphi(t)|\sigma_z|\varphi(t)\rangle$ has been calculated, and results are
plotted in Fig.~\ref{dynamics} for $\varepsilon=0.1$ (upper panel)
and $\sqrt{0.5}$ (lower panel) for coupling strengths
$g=0.1,0.2,0.5$. For comparison, results from exact
diagonalization and those of VVP are also shown.
It is found that the time-dependent
analytical results agree well with the numerical
ones, with substantial improvements over those obtained by VVP. It follows that the
contribution of the counter-rotating interaction is well taken in account in the BGRWA analytical solution.
Thus, our BGRWA approach is valid in a wide range of
coupling strengths for dynamic simulation of the wave functions.

\section{conclusion}

Analytical expressions without the RWA have been derived for the energy spectrum of the qubit-oscillator system with a finite bias.
Our approach takes into account the counter-rotating interactions while retaining mathematical simplicity of the ordinary RWA.
Eigenvectors and eigenvalues obtained analytically recover the results of GRWA at zero bias, and the BGRWA approximation is valid even in the ultrastrong coupling regime.
Our analytical spectrum expressions are shown in good agreement with experiment, and in comparison with
energy levels calculated using the VVP method and exact diagonalization, exhibit a wide range of validity for coupling strengths $g<0.5$.
Energy levels of the ground and lower-lying excited states obtained in this work show substantial improvements over the VVP results. In particular, our analytical expressions for the energy levels fit well with exact diagonalization results in the positive detuning regime, where the VVP method is invalid. Moreover, time evolution of $\sigma_z(t)$ obtained using BGRWA is in quantitative agreement with the exact diagonalization result for weak and ultra-strong couplings. The analytical approach presented here can be easily implemented to simulate
superconducting qubit-oscillator systems for coupling strengths up to $g=0.5$.
Finally, our approach can be employed to tackle problems of higher complicity such as a biased spin-boson model.

\acknowledgments

This work was supported by National Basic
Research Program of China (Grant Nos.~2011CBA00103 and
2009CB929104), National Natural Science Foundation of
China (Grants No.~11174254 and No.~11104363), and Research Fund for the
Doctoral Program of Higher Education of China (20110191120046).



\begin{references}
\bibitem{Scully} M. O. Scully and M. S. Zubairy, Quantum  Optics, Cambridge University Press, Cambridge,
1997;   M. Orszag, Quantum Optics: Including Noise Reduction,Trapped
Ions, Quantum Trajectories, and Decoherence, Science publish,
(2007); D. F. Walls and G. J. Milburn, Quantum Optics (Springer Verlag, Berlin, 1994).
\bibitem{Buluta} I.~Buluta et al.,  Science \textbf{326}, 108 (2009); Reports on Progress in Physics \textbf{74},104401 (2011).
\bibitem{schuster} D. I. Schuster et al., Nature (London) \textbf{445}, 515(2007)
\bibitem{Wallraff} A. Wallraff et al., Nature (London)\textbf{431}, 162(2004).
\bibitem{chio} I. Chiorescu, P. Bertet, K. Semba, Y. Nakamura, C. J. P. M. Harmans, and J. E. Mooij, Nature (London) \textbf{431}, 159(2004).
\bibitem{you} J.Q.~You et al., Phys. Today \textbf{58}, 42 (2009);  Nature \textbf{474}, 589 (2011);  Phys. Rev. B. \textbf{68}, 024510 (2003);  Phys. Rev. B. \textbf{68}, 064509 (2003).

\bibitem{nation} P.D.~Nation et al., Rev. Mod. Phys. \textbf{84}, 1 (2012).
\bibitem{jaynes} E.T.~Jaynes, and F.W.~Cummings, Proc. IEEE. \textbf{51}, 89(1963).

\bibitem{Niemczyk} T. Niemczyk et al., Nature Physics \textbf{6}, 772(2010).
\bibitem{pfd}   P. Forn-D\'{i}az et al., Phys. Rev. Lett. 105, 237001 (2010).
\bibitem{fedorov}  A. Fedorov et al., Phys. Rev. Lett. 105, 060503 (2010).
\bibitem{Ashhab}   S. Ashhab and F. Nori, Phys. Rev. A 81, 042311 (2010).
\bibitem{grifoni} J. Hausinger and M. Grifoni, Phys. Rev. A.
\textbf{82}, 062320 (2010).
\bibitem{irish} E.K.~Irish, Phys. Rev. Lett. \textbf{99}, 173601(2007).
\bibitem{feranchuk} I D Feranchuk, L I Komarov, and A P Ulyanenkov, J. Phys. A: Math. Gen. 29, 4035 (1996).
\bibitem{amniat} M Amniat-Talab, S Guerin, and H R Jauslin, J. Math. Phys. 46, 042311 (2005).

\bibitem{casanova} J. Casanova, G. Romero, I. Lizuain, J. J. Garcia-Ripoll, and E. Solano, Phys. Rev. Lett. \textbf{105},
263603(2010).
\bibitem{zhang} Y. Y. Zhang, Q. H. Chen, and S. Y. Zhu,  arXiv:1106.2191 (2011).

\bibitem{yu} L. X. Yu, S. Q. Zhu, Q. F. Liang, G. Chen, and S. T. Jia, Phys. Rev. A \textbf{86}, 015803 (2012).
\bibitem{zheng} C. J. Gan, and H. Zheng, Eur. Phys. J. D {\bf 59},473 (2010).
\bibitem{talab} M. Amniat-Talab et al., J. Math. Phys. \textbf{46}, 042311(2005).

\bibitem{lei}   Q.H.~Chen, L.~Li, T.~Liu, and K.L.~Wang, Chinese~Phys.~Lett.~{\bf 29}, 014208 (2012).


\end{references}
\end{document}